%% file: main.tex
\documentclass[runningheads]{llncs}

\usepackage[T1]{fontenc}
\usepackage{tikz}
\usepackage{amsmath}
\usepackage[]{graphicx}
\graphicspath{{./images/}}
\usepackage{algorithm}
\usepackage[noend]{algpseudocode}
\usepackage{booktabs}
\usepackage{caption}
\usepackage{multirow}
\usepackage{multicol}
\usepackage{makecell}
\usepackage{tabularx}
\usepackage{array}
\usepackage{threeparttable}
\usepackage{xcolor}
\usepackage{enumitem}
\usepackage{url}
\usepackage{hyperref}
\usepackage{listings}
\usepackage[algo2e]{algorithm2e}
\usepackage{float}
\usepackage[utf8]{inputenc}
\usepackage[framemethod=tikz]{mdframed}
\usepackage{relsize}
\usepackage{xspace}
\usepackage{ulem}

\definecolor{codebg}{gray}{0.97}
\definecolor{keywordcolor}{rgb}{0.2,0.2,0.7}
\definecolor{commentcolor}{gray}{0.45}
\definecolor{stringcolor}{rgb}{0.2,0.5,0.3}
\definecolor{codegreen}{rgb}{0,0.6,0}
\definecolor{codegray}{rgb}{0.5,0.5,0.5}

\lstdefinelanguage{Rust}{
    keywords=[1]{as,break,const,continue,crate,else,enum,extern,false,fn,for,if,impl,in,let,loop,match,mod,move,mut,pub,ref,return,self,static,struct,super,trait,true,type,unsafe,use,where,while,async,await,dyn},
    keywords=[2]{Self,Copy,Send,Sized,Sync,Drop,Fn,FnMut,FnOnce,Box,Vec,String,Option,Result,Some,None,Ok,Err,From,Into,Default},
    keywordstyle=[1]\color{codegreen}\bfseries,
    keywordstyle=[2]\color{blue}\bfseries,
    commentstyle=\color{codegray},
    stringstyle=\color{red},
    morecomment=[l][\color{gray}]{//},
    morecomment=[s][\color{gray}]{/*}{*/},
    morestring=[b]{"},
}

\lstdefinestyle{lst}{
    language=Rust,
    frame=lines,
    framesep=2mm,
    basicstyle=\ttfamily\footnotesize,
    columns=fixed,
    showstringspaces=false,
    numbersep=3pt,
    escapeinside={||},
    numbers=left,
    stepnumber=1,
    breaklines=true,
    breakatwhitespace=true,
    captionpos=b,
    float=t
}

\newcommand{\bheading}[1]{{\noindent{\textbf{#1}}}}

\newcommand{\rqone}{{Correctness of Compilation}\xspace}

\newcommand{\rqtwo}{{Effectiveness of Safety Enhancement}\xspace}

\newcommand{\rqthree}{{Introduction of Additional Rust Bugs}\xspace}

\newcommand{\cc}[1]{\mbox{\smaller[0.5]\texttt{#1}}}

\newcommand{\ru}{$Rust_{unsafe}$}
\newcommand{\ridio}{$Rust_{idiomatic}$}
\newcommand{\ana}{C2Rust-analyze\xspace}
\newcommand{\crown}{CROWN\xspace}
\newcommand{\safer}{C2SaferRust\xspace}
\newcommand{\fluo}{FLUORINE\xspace}

\newcommand{\raf}[1]{\textcolor{black}{#1}}

\def\Snospace~{\S{}}

\renewcommand{\figurename}{Fig.}

\begin{document}

\title{C-to-Rust Fallacy: Automatic Refactoring $\neq$ Memory Security}
\titlerunning{C-to-Rust Fallacy}
\author{Hung-Mao Chen\inst{1} \and
Xu He\inst{2} \and
Bo Lu\inst{1} \and
Xiaokuan Zhang\inst{1} \and
Kun Sun\inst{1}}

\authorrunning{H.-M. Chen et al.}

\institute{George Mason University\\
\email{\{hchen28,blu5,ksun3\}@gmu.edu}, \email{xiaokuan.zhang.cs@gmail.com} \and
Visa Inc.\\
\email{xuhe3@visa.com}}

\maketitle

\begin{abstract}
Rust has emerged as the leading system programming language, offering strong memory and type safety guarantees without compromising performance. This positions it as a compelling alternative to traditional languages like C and C++, which are susceptible to memory security bugs. However, manually transforming C to Rust requires in-depth domain knowledge of the Rust language features, which requires significant effort for developers. To address this, tools for automatic C-to-Rust refactoring aim to generate safe Rust code leveraging static analysis and Large Language Models (LLMs). While these tools claim to achieve safety by reducing the unsafe Rust, the correlation with improving security is not clear. In this paper, we conduct a comprehensive empirical study on the reliability, safety, and correctness of various C-to-Rust refactoring methods. Specifically, we evaluate C2Rust-analyze, CROWN, C2SaferRust, and FLOURINE using a dataset of 116 C programs with memory security bugs from the NIST Juliet Test Suite. Based on 464 Rust programs generated by these tools, our evaluation focuses on three key aspects: the compilation correctness of the refactored programs, the effectiveness in mitigating original C bugs, and the tendency to introduce additional Rust bugs. The results indicate that \raf{342} Rust programs fail to compile, \raf{177 Rust programs inherit memory security bugs from the original C programs}, and \raf{77} new Rust bugs are introduced. We examine the rationale behind tool design and analyze the root cause of errors across various refactoring methods. Our findings indicate that current automated refactoring tools deliver memory safety as they define it, but not the broader memory security when adopting them.
\end{abstract}


\input{Introduction}
\input{Background}
\input{Dataset}
\input{RQ1}
\input{RQ2}
\input{RQ3}
\input{Discussion}
\input{RelatedWork}
\input{Conclusion}

\appendix

\input{Appendix}

\bibliographystyle{splncs04}
\bibliography{reference}

\end{document}

%% file: Introduction.tex
\section{Introduction}

Rust~\cite{RustLang} is a modern programming language renowned for its stringent enforcement of type safety and memory security through compile-time checking, while maintaining high runtime performance. 
These features make Rust an attractive alternative for programs written in C and C++, which are prone to memory security issues such as Null Pointer Dereference (NPD), Use-After-Free (UAF), and Buffer Overflow (BOF).
Rust has been gradually adopted in major C projects such as the Linux kernel~\cite{LinuxKernel}, Windows~\cite{windows}, and the Firefox browser~\cite{MozillaFirefox}. 
Furthermore, the White House has advocated for transitioning to memory-safe programming languages, such as Rust, to enhance cybersecurity across systems~\cite{whitehouse}.

Rust can be divided into two parts: 1) \cc{safe} Rust, which enforces strict compile-time checks to maintain memory security and type safety; and 2) \cc{unsafe} Rust, which relaxes the safety check to allow risky operations (e.g., accessing raw pointers), but at the same time voids the safety guarantees provided by Rust and can lead to security issues~\cite{Xu2020MemorySafetyCC,Yechan2021Rudra,Zhuohua2021MirChecker}.
Converting C code into Rust manually necessitates a comprehensive grasp of Rust's features, presenting a significant learning challenge for developers. Additionally, the resultant Rust code frequently includes \cc{unsafe} Rust code, undermining Rust's inherent safety guarantees.

To aid the transition from C to Rust and maximize the usage of safe Rust code, automatic tools from both industry and academia have been developed to 
\raf{leverage Rust's type system for safety enhancement.}
These tools can be classified into two categories.
1) \textbf{C2Rust-based refactoring} (e.g., CROWN~\cite{crown}): These tools first use C2Rust to convert C code to unsafe Rust (translation process), then leverage static analysis or Large Language Models (LLMs) to refactor unsafe Rust code into safe Rust code (refactoring process)~\cite{laertes,crown,hong2024don,nitin2025c2saferrust}.
2) \textbf{Direct refactoring} (e.g., FLUORINE~\cite{fluorine}): These tools directly perform translation and refactoring processes in one step, usually leveraging LLMs~\cite{yang2024vert,shetty2024syzygy}.
Although these tools \raf{claim to enhance the safety} and achieve some success, their primary focus is on minimizing the usage of unsafe code, often neglecting other crucial aspects, such as the security impacts. 
Previous research~\cite{userstudy} has primarily assessed C-to-safe-Rust conversion techniques by examining the proportion of safe code produced and the likelihood that the refactoring passes test suites.
However, it does not answer the security-relevant question: if the input C program contains a memory-safety vulnerability, whether the refactored Rust output eliminates it, preserves it so that developers can audit, or silently alters program behavior while introducing new bugs of its own.
While these refactoring tools claim to achieve safety, their correlation with security impact is not clear.
In other words, the implicit assumption that reducing \cc{unsafe} surface correlates with improved security has never been validated.

In this paper, we conduct the security-focused evaluation of automated C-to-Rust refactoring, examining whether state-of-the-art tools can deliver the memory security in the output program. 
We do not assume this is the only reasonable goal for a refactoring tool, but we evaluate the security impact that can be brought by these tools.
First, we collect a dataset of 116 C programs from the NIST Juliet Test Suite~\cite{juliet-test-suite}, which is widely used in research studies~\cite{julietone,juliettwo}.
Specifically, C programs in our dataset include the bugs of Null Pointer Dereference (NPD), Use-After-Free (UAF), Double Free (DF), Buffer Overflow (BOF), which are among the top 25 bug types most exploited in the CWE list for 2025~\cite{top25}, and Type Confusion (TC).
These buggy programs are also the best cases to demonstrate Rust's superiority in memory security, since safe Rust is supposed to prevent these bugs with designs of ownership and type system. \raf{For example, while the original C program re-interprets a character type as an integer type, and leads to a type confusion bug, the same type conversion behavior in Rust program is forbidden by the Rust compiler and will trigger syntax error during the compilation phase.} 
To cover as various types of bug scenario as possible, we pick two C programs for each bug features in the Juliet Test Suite (58 different bug features in total).
These features consists of different type of pointers and different functions or operations that trigger the bugs.
Based on the dataset, we apply three C2Rust-based tools (C2Rust-analyze~\cite{c2rustanalyze}, CROWN~\cite{crown}, C2SaferRust~\cite{nitin2025c2saferrust}) and one direct tool (FLUORINE~\cite{fluorine}) to generate 464 new Rust programs. These tools, published in the last two years, demonstrate state-of-the-art refactoring techniques for generating idiomatic Rust code.
Based on these newly generated codes, we investigate three research questions about the refactoring process.

\bheading{RQ1: \rqone.} The initial research question is to evaluate \raf{whether the resulting Rust code can be compiled.} In this experiment, we assess this RQ by observing whether the output messages of \cc{cargo build} include any errors. Furthermore, an LLM agent is set up to automatically fix compilation errors, which facilitates the investigation on following research questions.

\bheading{RQ2: \rqtwo.}
Second, we evaluate whether automatic refactoring can retain the memory security capabilities of safe Rust. \raf{We assess the effectiveness of safety enhancement by verifying the pre-existing C bugs can be fixed in the resulting Rust code}.
Since Address Sanitizer (ASan~\cite{serebryany2012addresssanitizer}) is effective in identifying most memory security issues, we utilize it to confirm that existing bugs have been resolved by comparing the backtrace. 
However, for certain bugs that are only triggered in specific scenarios (e.g., a null pointer caused by out-of-memory conditions), we rely on manual verification to ascertain the results.

\bheading{RQ3: \rqthree.} 
In addition to original C bugs, we examine the possibility that these refactoring tools aggressively introduce new bugs or runtime security checks that do not exist in original C code. 
To address RQ3, we run static analyzers such as Clippy~\cite{rustclippy} to detect potential issues introduced by refactoring. To expand the scope of static analysis, we also set up an LLM agent guided by rules of undefined behaviors in Rust and interactively using ASan and Miri~\cite{miri} for verification.

For each research question, we summarize the results and perform root-cause analysis of errors by investigating the corresponding refactoring strategies.
For RQ1, the refactoring techniques that apply LLMs (FLUORINE) achieve the highest compilation correctness.
Pure static analysis methods are prone to compilation errors due to strategies such as incremental rewriting or unstable features.
For RQ2, our study shows that there are also \raf{177} original memory security bugs that are retained due to syntax refactoring and precision-based type refactoring.
For RQ3, our findings indicate that these tools introduce 77 additional bugs caused by translation and refactoring, such as run-time panic across FFI boundaries.

Our study makes the following contributions:
\begin{itemize}
    \item We show that automated C-to-Rust refactoring fails to deliver the memory-security guarantees, which means tools' users should still audit the output Rust programs.

    \item We expose new attack interfaces, e.g., panic across FFI boundaries, can happen in the scenario of automated refactoring.

    \item We connect concrete refactoring design decisions, e.g., precision-based rollback, LLM prompt constraint conflicts, to the specific security failure modes they produce.
    
    \item We release our datasets and test scripts to enable reproducibility and future benchmarking of security-oriented C-to-Rust refactoring.
\end{itemize}

%% file: Background.tex
\section{Background}

\subsection{C-to-Rust Refactoring Tools}

\begin{figure}
    \centering
    \includegraphics[width=0.8\linewidth]{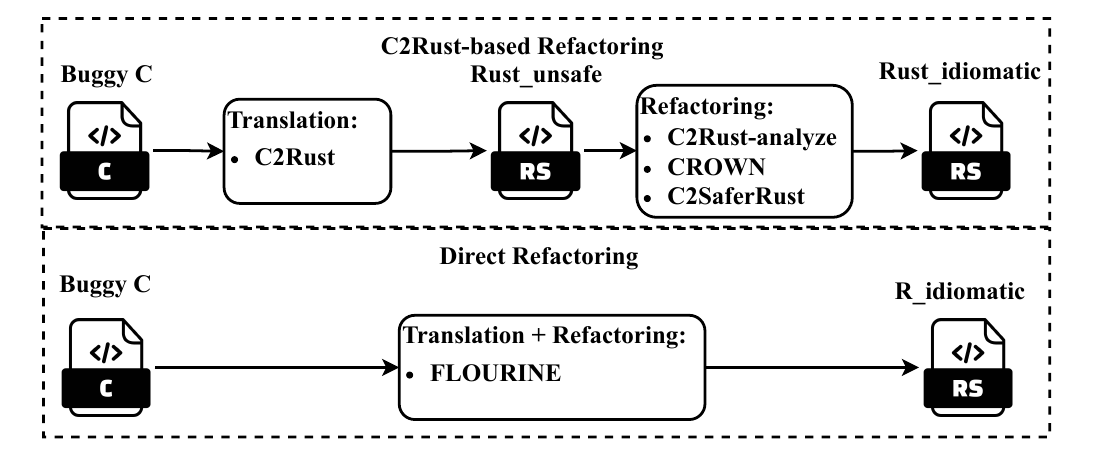}
    \caption{Two categories of automatic C-to-Rust refactoring techniques: C2Rust-based refactoring consists of translation (output: \ru) and refactoring processes (output: \ridio) while direct refactoring completes them in one step (output: \ridio).}
    \label{fig:c2rust}
\end{figure}

Automatic C-to-Rust refactoring tools aim to generate idiomatic Rust code. The key principles are: leverage Rust's ownership system, handle errors explicitly, prefer iterators, use pattern matching extensively, and let the type system prevent bugs.
Based on the different stages in the refactoring process, we categorize these tools into two types: \textit{C2Rust-based} and \textit{Direct} refactoring (see~\autoref{fig:c2rust}).

\bheading{C2Rust-based refactoring.} 
This approach consists of two separate processes: translation and refactoring.
In the translation process, C2Rust~\cite{c2rust} analyzes the abstract syntax tree (AST) of the C code to generate syntactically equivalent unsafe Rust code (\ru). The generated code still contains many unsafe Rust snippets and residual C code fragments because many pointer operations in C programs cannot be directly translated into safe Rust code.
In the next step, refactoring tools attempt to convert unsafe code into safe Rust, creating idiomatic Rust code (\ridio).
Note that refactoring tools usually cannot completely remove all unsafe code.
Refactoring tools analyze whether programs contain predefined code patterns, and then infer safety semantics such as lifetime or ownership. 
CROWN~\cite{crown} analyzes Rust programs and generates ownership constraints, then refactors unsafe pointers when constraints can be solved by the SAT solver.
While Laertes~\cite{laertes} and CROWN both focus on unsafe pointer refactoring, Nopcrat~\cite{hong2024don} and Urcrat~\cite{10.1145/3691620.3694985} focus on refactoring algebraic data types and union types.
In addition to the efforts made by the Rust community, the C2Rust team releases their own refactoring plugin, C2Rust-analyze~\cite{c2rustanalyze}.
C2Rust-analyze applies a novel methodology, \textit{permission analysis} to assess the required permissions for different memory operations (e.g., \cc{WRITE}, \cc{FREE}, \cc{OFFSET\_ADD}).
Based on the permission set, the tool can refactor pointers into suitable types.
In addition to static analysis, tools can also use LLMs to generate idiomatic Rust code during the refactoring process.
C2SaferRust~\cite{nitin2025c2saferrust} uses static analysis (dataflow and callgraph analysis) to convert unsafe Rust programs into code slices, then provides this information to LLMs for safe code refactoring.

\bheading{Direct refactoring.}
This approach performs translation and refactoring processes in a single step.
Most tools in this category use LLMs to refactor C code into idiomatic Rust code.
FLUORINE~\cite{fluorine} translates C/Go functions to Rust and verifies the equivalence of I/O through differential fuzzing between languages.
VERT~\cite{yang2024vert} can translate C/Go and also C++ functions, but builds a trusted oracle via WebAssembly and verifies semantic equivalence using formal verification.
Similar to C2SaferRust, Syzygy~\cite{shetty2024syzygy} also uses static analysis to create code slices and provide information to LLMs.
However, Syzygy integrates dynamic analysis on test cases, which are generated by LLMs for I/O equivalence.
If the refactoring fails the runtime tests, it returns error feedback and retrains.

\subsection{Rust and Memory Security Bugs}

Safe Rust ensures memory security at compile time through its ownership and borrowing model. Every value in Rust has a single variable as its owner. When the owner goes out of scope, the memory allocated for the value is automatically released, and any subsequent access to the variable triggers compilation errors. This prevents classic memory security bugs such as \textit{double free} and \textit{use-after-free} from accessing illegal memory. For \textit{null pointer dereference}, safe Rust provides the \cc{Option} type to represent nullability. This approach forces developers to explicitly handle cases where values are not present, thereby preventing null pointer dereference. For \textit{buffer overflow} and \textit{type confusion}, the compiler inserts length checks and explicit type conversion rules to mitigate bugs at both compile-time and run-time.

However, there are certain low-level operations that are impossible for a compiler to statically check, e.g., dereferencing raw pointers, calling unsafe functions, accessing mutable static variables, implementing unsafe traits, and writing inline assembly~\cite{unsafeRu19online}.
The \cc{unsafe} keyword is required to mark this code, which requires developers to ensure safety themselves. One common case where \cc{unsafe} is required is the implementation of the \cc{Vec<T>} type in safe Rust. In the standard library, it is actually implemented by managing raw pointers for memory allocation, though its public API is marked as safe. The presence of unsafe Rust also provides opportunities for memory security bugs to escape the compile-time and run-time checks of the Rust compiler.

%% file: Dataset.tex
\section{Overview}

\begin{table}[t]
\centering
\caption{C-To-Rust tool list (By May 1, 2025).}
\label{tab:tool-selection}
\footnotesize
\begin{tabular}{l|c|c|c}
\toprule
\textbf{Tools} & \textbf{Source Code} & \textbf{Scope} & \textbf{Analysis} \\
\midrule
\textbf{CROWN}~\cite{crown} & Available & \raf{Pointer} & Program Analysis \\
\textbf{C2Rust-analyze}~\cite{c2rustanalyze} & Available & \raf{Pointer} & Program Analysis \\
\textbf{FLUORINE}~\cite{fluorine} & Available & General & LLM \\
\textbf{C2SaferRust}~\cite{nitin2025c2saferrust} & Available & General & Program Analysis + LLM \\
Nopcrat~\cite{hong2024don} & Available & ADT & Program Analysis \\
Urcrat~\cite{10.1145/3691620.3694985} & Available & Union & Program Analysis \\
Syzygy~\cite{shetty2024syzygy} & Not Available & General & Program Analysis + LLM \\
VERT~\cite{yang2024vert} & Available & General & LLM \\
\bottomrule
\end{tabular}
\end{table}

\subsection{Tool Selection}

We choose automated refactoring tools from the eight targets that are released between Jan 1, 2023 and May 1, 2025 (see~\autoref{tab:tool-selection}). 
Among the seven tools, four of them apply static analysis while the others apply LLMs for refactoring techniques.
All tools aim to generate idiomatic Rust code.
Nopcrat's and Urcrat's refactoring especially focus on specific data structure, but we aim at general and complementary types.
Therefore, we exclude them.
Syzygy is not chosen because it is not open-source.
While VERT provides source code and compatible refactoring scope, it requires users to follow the specific syntax requirements in C code that could break the consistency of our dataset.
Therefore, we select CROWN, C2Rust-analyze, FLUORINE, and C2SaferRust for our evaluation.

\subsection{Data Collection}

We derive our dataset from the top 25 known exploited bugs in the CWE catalog for 2025~\cite{top25}, focusing on following memory-related bugs: \textit{Null Pointer Dereference (NPD)}, \textit{Use After Free (UAF)}, \textit{Double Free (DF)}, \textit{Buffer Overflow (BOF)}, then supplement with \textit{Type Confusion (TC)}.
We exclude the remaining bugs from the top 25 list because they represent web-based security bugs and logic-based security bugs that cannot be addressed through C-to-Rust code refactoring.
Our dataset comprises 116 representative C programs sourced from the \textit{NIST Juliet Test Suite} \cite{juliet-test-suite}.
The selected samples encompass CWE690 and CWE476 for null pointer dereference, CWE416 for use-after-free, CWE415 for double-free, CWE121 and CWE122 for buffer overflow, and CWE843 for type confusion (see~\autoref{app:bugtype}).
To ensure comprehensive coverage of various scenarios, we select C programs according to specific features provided by the NIST Juliet Test Suite.
Within each CWE category, we select programs to cover different trigger patterns (e.g., allocation methods for NPD, allocation locations for BOF, released pointer types for UAF/DF/memory access patterns). Detailed feature coverage is listed in~\autoref{app:datafeature}.
We distribute the samples across each memory security bug category to ensure our dataset encompasses all relevant bug scenarios available in the NIST Juliet Test Suite.

\subsection{Research Questions and Workflow}

\begin{figure}[t]
    \centering
    \includegraphics[width=0.8\linewidth]{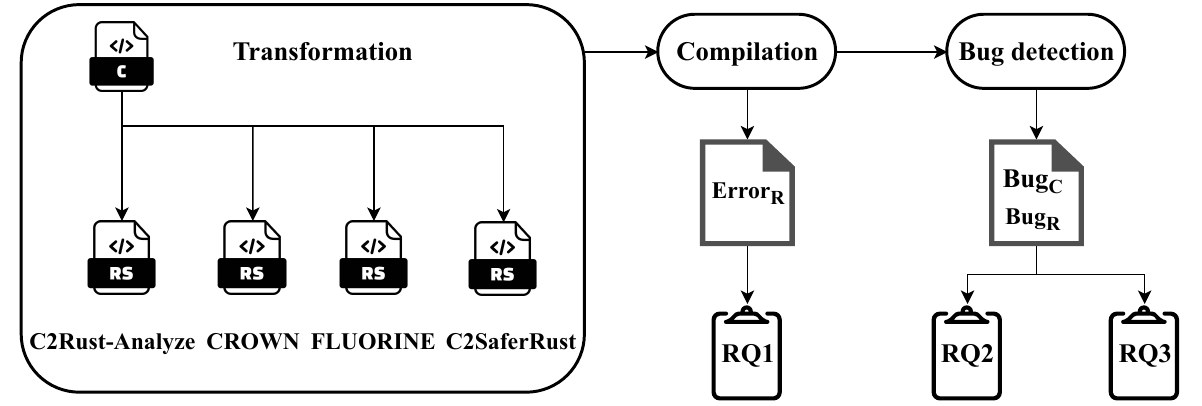}
    \caption{The workflow of our study. Inputs are a C program and existing bugs in that program ($Bug_C$). After applying refactoring, we obtain Rust code $R_{idiomatic}$. We then perform compilation, bug detection to get compilation error ($Error_R$), bugs ($Bug_C$, $Bug_R$).}
    \label{fig:workflow}
\end{figure}

Based on the selected tools and dataset, we investigate the challenges and implications of automatic C-to-Rust refactoring. To systematically assess the reliability and security issues of current refactoring tools, we apply them to C code with existing bugs $Bug_C$, then generate the Rust code ($R_{idiomatic}$). Based on C code and refactored Rust code, we formulate three research questions and their workflow as in~\autoref{fig:workflow}.

\bheading{RQ1: Do refactoring tools generate correctly compiled Rust code?}
We check whether existing tools can successfully generate compilable Rust programs. Given the substantial differences in syntax and type semantics between C and Rust, compilation failures are common in refactored code. Understanding whether and why such failures occur provides insight into the maturity and robustness of these tools.
First, we compile the program $R_{idiomatic}$ to obtain the compilation errors ($Error_R$). We check if $Error_R$ exists and study the root cause.

\bheading{\raf{RQ2: Do refactoring tools inherently enhance the safety of Rust code?}}
We focus on whether the refactoring process can eliminate existing C bugs ($Bug_C$), which include five types of memory security bugs from our dataset. Rust's built-in types and automatic memory management can resolve memory-safety bugs to a certain degree, especially when applying refactoring or strict compliance with Rust's ownership model. We further analyze the root causes of any remaining bugs after refactoring.
We run ASan on $R_{idiomatic}$ in the bug detection phase to obtain the results ($Bug_R$). We then check if $Bug_C$ overlaps with $Bug_R$ and study the root cause of residential $Bug_C$.

\bheading{RQ3: Do refactoring tools introduce additional Rust bugs?}
We examine whether refactoring tools introduce additional Rust bugs, which could be Rust's undefined behaviors or extra security checks that do not exist in C code. Many refactoring tools, especially those that utilize LLMs, may substantially rewrite logic or restructure program control flows. These changes increase the potential for new bugs, particularly in scenarios where unsafe Rust code blocks are still present. We run static analyzers such as Clippy and an augmented LLM agent on $R_{idio}$ in the bug detection stage to obtain the results ($Bug_R$). The LLM agent uses Miri to help verify that $Bug_R$ exists. Within $Bug_R$, we exclude existing $Bug_C$ if it exists, then summarize the root causes of additional bugs.

%% file: RQ1.tex
\section{RQ1: \rqone}

In this section, we evaluate the compilation correctness of Rust programs generated by four refactoring tools. We first collect and categorize the compilation errors encountered in the refactored code. Next, we analyze the underlying causes by examining the details of the implementation of each tool. Since LLM-based error fixing is increasingly adopted in practice, we apply the same process to refactoring tools, which still rely on static analysis alone. In the end, we summarize the refactoring strategies that enhance compilation correctness and their impact on failures.

\begin{table}[t]
\centering
\caption{Compilation availability of C-To-Rust transformation. Note that, for compilation failures ($x|y$), $x$ and $y$ refer to the count of compilation failures before and after adopting LLMs for fixing compilation errors, respectively.
Error Codes ($E*$ or multiple ones splitted with $/$) are emitted by the Rust compiler according to Rust Error Codes Index~\cite{Rust-error-code}.
}
\footnotesize
\begin{tabular}{c c c c c}
 \toprule
 {} & \multicolumn{3}{c}{\textbf{C2Rust-based}} & \textbf{Direct}  \\ \cmidrule{2-5}
 \textbf{\begin{tabular}[c]{@{}c@{}}Compilation\end{tabular}} & \textbf{\begin{tabular}[c]{@{}c@{}}C2Rust-analyze\end{tabular}} & \textbf{CROWN} & \textbf{\begin{tabular}[c]{@{}c@{}}C2SaferRust\end{tabular}} & \textbf{\begin{tabular}[c]{@{}c@{}}FLUORINE\end{tabular}} \\
 \midrule
 Compilation Failures & 98 | 29 & 116 | 11 & 116 | 11 & 12 | 0  \\
 \midrule
 Error Codes & \begin{tabular}[c]{@{}c@{}}E0308\\E0605\\E0658\end{tabular} & \begin{tabular}[c]{@{}c@{}}E0658/E0433\\E0614/E0308\\E0745/E0277\end{tabular} & \begin{tabular}[c]{@{}c@{}}E0308\\E0133\\E0433\end{tabular} & \begin{tabular}[c]{@{}c@{}}E0308\\E0599\\E0412/E0433\end{tabular} \\
 \bottomrule
\end{tabular}
\label{tab:compilability}
\end{table}

\subsection{Methodology}
We first run \cc{cargo build} on the refactored Rust code.
If there is an error, it indicates that the code cannot be correctly compiled, then we study the root causes of compilation errors.
If there is no error, we also study whether any errors are automatically fixed in the refactoring process.

We find that there are a large number of errors that can be fixed by small code changes (e.g., import compiler features). Therefore, we implement an LLM agent to perform trivial fixes for us.
\raf{The agent is powered by \textit{Claude Sonnet 4}~\cite{claude4}, which is known with excellent coding capability.}
The agent design is to use \cc{cargo build} as a function tool and fix compilation errors based on the returned messages in five attempts. The purpose is to summarize compilation errors that are not resolved by the LLM. Furthermore, the correctly compiled Rust code can help us move forward to next research questions on more targets.

\subsection{Results}

As shown in~\autoref{tab:compilability}, all refactoring tools produce programs with compilation errors before applying the LLM error-fixing mechanism.
Type mismatch errors (E0308) are the most prevalent issue, appearing in Rust programs generated by all refactoring tools. Additionally, unstable library feature errors (E0658) occur in all programs generated by \crown.

Our experiments with automatic error fixing demonstrate the LLM's effectiveness in error correction and substantial improvements in compilation pass rates across all tools.
The LLM agent successfully applies targeted fixes, including adding \cc{as\_ptr} methods to resolve invalid cast errors (E0605) when converting slices to raw pointers. For complex wrapper types involving \texttt{Option}, the agent corrects improper dereference operations by replacing them with \cc{unwrap} calls. When encountering unstable Rust features, the agent addresses these issues by incorporating the necessary feature flags and external crate dependencies.
Our evaluation of \safer reveals that while LLMs can initially generate programs with numerous compilation errors, the multi-round debugging mechanism achieves a 94\% improvement within 4 iterations.

\subsection{Root Cause Analysis}

Based on our analysis of common compilation errors, we conduct a root cause investigation to identify the refactoring strategies most responsible for these issues. We examine three error-prone strategies, their underlying rationale, and their connection to specific compilation failures.

\begin{lstlisting}[
language=rust,
style=lst,
caption={C2Rust-analyze converts \cc{callee} function with type refactoring (see line 3) and keep the \cc{caller} function remained. It generates \cc{callee\_shim} to bridge the call chain between two functions (see line 7).},
label=eliunsafe,
mathescape=true,
float=t,
basicstyle=\ttfamily\scriptsize,
belowskip=-0.5em
]
pub unsafe extern "C" fn callee<'h0,'h1>(mut dataPtr: &'h0 (&'h1 mut [libc::c_int])) {  
    // Expects: &'h0 (&'h1 mut [libc::c_int]) 
    let mut data: &mut [libc::c_int] = *dataPtr;
    let mut source: [libc::c_int; 10] = [0 as libc::c_int, 0, 0, 0, 0, 0, 0, 0, 0, 0];
}    // ...

unsafe fn callee_shim(arg0: *mut *mut libc::c_int) {
    let safe_arg0 = &*arg0.cast_const(); // Generates: &*mut i32
    let safe_result = callee(safe_arg0);
} // ...

pub unsafe extern "C" fn caller() {
    let mut data: *mut libc::c_int = 0 as *mut libc::c_int;
    data = 0 as *mut libc::c_int;
    let mut fresh0 = std::vec::from_elem(0, 10 as libc::c_int as libc::c_ulong as usize);
    data = fresh0.as_mut_ptr() as *mut libc::c_int;
    callee_shim(*mut data);
}
\end{lstlisting}

\bheading{Type refactoring creates type mismatches between refactored local code and unmodified globals.}
This strategy converts unsafe pointers into safe types that enable compiler ownership tracking. The refactoring replaces C pointers with safer Rust abstractions such as \cc{\&[T]}, \cc{\&mut [T]}, \cc{Option<\&T>}, or \cc{Box<T>} according to specific C code patterns.
\ana employs this strategy by independently tracking and refactoring each pointer within individual function scopes.
For instance, \ana refactors local pointers (e.g., \cc{*mut T} to \cc{Option<\&T>}) while intentionally preserving global variables and static data in their original types.
This design aims to enhance memory security while maintaining interface compatibility across the entire code base.
However, this approach creates a problematic boundary between refactored local code and unmodified global state, resulting in type mismatch errors (E0308) and invalid cast errors (E0605).
Consider a scenario where a local pointer \cc{data} is assigned \cc{NULL} in C code and later assigned to a global pointer \cc{static\_data}.
While \ana safely converts \cc{data} into a reference type wrapped in an \cc{Option}, \cc{static\_data} remains declared as a raw pointer type.
This demonstrates that independent type refactoring introduces persistent type mismatch errors that remain unresolved even after applying LLM's error-fixing.

\bheading{Incremental rewriting fails to handle complex type conversions between safe and unsafe interfaces.}
Unlike comprehensive ownership tracking approaches, this method enables a gradual transition of large C code bases without requiring complete rewrites. The real-world C programs contain a mixture of code where some pointers can be refactored into reference types in Rust, others require more complex analysis, or remain as raw pointer types. Rather than completely failing upon encountering unsafe code, this approach utilizes an incremental transition method that maximizes safety improvements while ensuring compatibility throughout all code bases. \ana implements this strategy through \textit{shim} function generation. Shim functions serve as adapters that bridge compatibility gaps between different interfaces. At interprocedural function boundaries, safe functions with refactored signatures use shims to communicate with legacy unsafe code. However, this design also introduces compilation errors. In the \cc{callee} function (see~\autoref{eliunsafe}), \ana identifies that the parameter \cc{dataPtr} can be safely refactored to a slice type (\cc{\&(\&mut [libc::c\_int])}) based on usage patterns such as dereferencing to obtain a mutable slice and copying data. Meanwhile, the caller function continues using raw pointer operations (\cc{*mut *mut libc::c\_int}). Our analysis reveals that \cc{callee\_shim} fails to properly handle complex type conversions from \cc{**T} to \cc{\&\&[T]}, resulting in type mismatch errors (E0308). These errors persist in the final results even after applying LLM's error-fixing processes.

\bheading{Reliance on compiler-specific features breaks compilation in production environments.}
This refactoring strategy utilizes features of the Rust compiler (\textit{rustc}) that require explicit opt-in via feature flags or toolchain configuration. The underlying rationale is to bridge specific gaps between C's low-level memory model and Rust's safety mechanisms by reaching into compiler internals or syntax extensions not part of standard Rust. \crown exemplifies this strategy by employing raw reference syntax, strict provenance methods, and core intrinsics, all of which require feature flags at compile time and trigger feature-gate errors (E0658) in environments without the corresponding configuration.
Consider raw references as an example: \crown automatically applies \cc{\&raw mut} to all mutable pointers during preprocessing.
During ownership analysis, \crown only refactors pointers that demonstrate ownership characteristics while preserving non-owning pointers with raw reference syntax.
Subsequent attempts to access these raw references result in field access errors (E0614). According to our dataset, \crown encounters these errors in all refactored Rust programs. \safer also shows these errors in the parts of programs that cannot be fixed by automatic process. This finding demonstrates that reliance on features requiring explicit compiler configuration can compromise compilation correctness in standard production environments.

\vspace{0.5em}
\begin{mdframed}[
  roundcorner=5pt,
  linewidth=0.8pt,
  linecolor=black,
  backgroundcolor=gray!20,
  innertopmargin=6pt,
  innerbottommargin=6pt,
  innerleftmargin=6pt,
  innerrightmargin=6pt
]
\textbf{Summary for RQ1.}
Refactoring strategies like type refactoring and incremental rewriting produce most compilation errors, driven by the need for compatibility with non-refactored code. In contrast, while the iterative fixing process of LLM-based approaches significantly increases compilation success, it may also include patterns that rely on feature flags or toolchain configuration to pass compilation checks.
\end{mdframed}

%% file: RQ2.tex
\section{RQ2: \rqtwo}
In this section,
we first exclude 51 (29 + 11 + 11 + 0) Rust programs that still have compilation errors after applying LLM's automatic error fixing in~\autoref{tab:compilability}.
In the remaining 413 Rust programs, we evaluate the effectiveness of safety enhancement by studying how refactoring can mitigate existing C bugs.
We run the bug detector on both the C and Rust code, then we manually compare the errors in the reports. In the end, we collect all remaining C bugs from Rust code and perform the root cause analysis for the failures.

\subsection{Methodology}
We first use ASan to verify whether existing C bugs can trigger the alarm. If it can be triggered, we can also run ASan on the refactored Rust code. After the alarm is raised, we manually compare the ASan error traces in both C and Rust codes to verify if it is the same root cause, indicating mitigation failures.
In other cases where ASan provides different error traces or empty outputs, we rely on separate criteria in different categories of bugs.
\raf{Since our targets are refactoring tools instead of translation tools, their outputs (Rust code) will introduce necessary changes~\cite{refactor-rust} by leveraging Rust's ownership-based model to enforce memory security. In the caption of~\autoref{tab:repair}, we summarize the corresponding strategies by which safe Rust can prevent the memory security bugs~\cite{rustwiki,typepulse,zhang2024beyond}. It helps us evaluate the security enhancements brought by refactoring.}
There are also the cases where the bug cannot be triggered. In the cases of CWE690 in our dataset, the bug occurs only when the programs fail to allocate the memory. In these cases, we manually inspect whether the refactored Rust code is also vulnerable to potential risks.

\begin{table}
\centering
\caption{Safety Enhancement in 413 Rust programs. $(I/II/III/IV)$ represents the refactoring strategies: I. \emph{Panic} (e.g., \texttt{unwrap} on \texttt{None}/\texttt{Err}, out-of-bounds); II. \emph{Error handling} (e.g., explicit \texttt{None}/null checks, bounds checks, assertions); III. \emph{Safe pointer types} (e.g., \texttt{Vec}, \texttt{Box}, references/slices); IV. \emph{Semantic fixes of illegal access} (e.g., remove use-after-free/drop, correct buffer sizes/type interpretation). The last row ($x/y$) shows the number of mitigated bugs and total programs.}
\footnotesize
\begin{tabular}{c c c c c}
 \toprule
 {} & \multicolumn{3}{c}{\textbf{C2Rust-based}} & \textbf{Direct}  \\ \cmidrule{2-5}
 \textbf{\begin{tabular}[c]{@{}c@{}}Bug Type\end{tabular}} & \textbf{\begin{tabular}[c]{@{}c@{}}C2Rust-analyze\end{tabular}} & \textbf{CROWN} & \textbf{\begin{tabular}[c]{@{}c@{}}C2SaferRust\end{tabular}} & \textbf{\begin{tabular}[c]{@{}c@{}}FLUORINE\end{tabular}} \\
 \midrule
 NPD (149) & 19 (I) & 0 & 35 (III, I, II) & 37 (III, II, I) \\
 UAF (54) & 8 (I) & 0 & 11 (III) & 14 (III, II, IV) \\
 DF (48) & 0 & 0 & 12 (III) & 12 (III, IV) \\
 BOF (150) & 12 (I, II, IV) & 4 (IV) & 23 (IV, II, I) & 42 (II, IV) \\
 TC (12) & 0 & 0 & 3 (IV) & 4 (IV) \\
 \midrule
 \textbf{\begin{tabular}[c]{@{}c@{}}Total (413)\end{tabular}} & 39 / 87 & 4 / 105 & 84 / 105 & 109 / 116 \\
 \bottomrule
\end{tabular}
\label{tab:repair}
\end{table}

\subsection{Results}
As illustrated in~\autoref{tab:repair}, the refactoring tools are able to address \raf{236} bugs (57.1\%) in 413 Rust programs that can be compiled.
Among the four refactoring tools, the LLM-based tools (\fluo and \safer)
improve the safety of the Rust programs more effectively by addressing 84 and 108 original bugs.
In contrast, static analysis-based tools (\ana and \crown) address fewer original bugs (39 and 4) although they also have successfully refactored some raw pointers.
Particularly within C code involving bug types such as DF and TC, static analysis tools do not address any of these issues, whereas LLM tools are capable of fixing nearly all of them. The results show that using LLM can significantly inherit the safety features of safe Rust and mitigate the existing C bugs while there are still several bugs remained in the refactored Rust programs.

Based on the 236 cases of mitigation, we then summarize 4 refactoring strategies in the caption of~\autoref{tab:repair}, including the safety enforced by the type system and the LLM's semantic refactoring, which are the major strategies that succeed in bug mitigation.
Safe Rust provides types such as reference, \cc{Vec}, and \cc{Box} types (III), which are tracked by ownership. We find that refactoring tools rely on these types to fix NPD, DF, and UAF bugs.
\raf{Rust type system also converts silent memory corruption in C into the runtime panic (I) with explicit diagnostics and controlled unwinding, preventing latent exploitation.}
In the compilation phase, the developers are required to conform to explicit type declaration and type conversion rules (IV), which fix BOF and TC bugs.
Based on the application of a type system mentioned above, the refactoring based on LLM first infers the semantics of C programs, then provides a more elegant refactoring. In the C code with BOF bugs, we observe that LLM will add bound check (II) instead of relying on runtime panic. There are 63 out of 236 mitigated bugs are added with such kinds of error handling, which also enhances the mechanism of error recovery for software development.

\subsection{Root Cause Analysis}
\label{miti-fail}

From our evaluation of the unresolved C bugs, we implement a root cause analysis to pinpoint the refactoring strategies responsible for these mitigation failures. We provide a summary of three error-prone strategies, their objectives, and how they relate to persistent bugs.

\begin{table}[t]
\centering
\caption{Root cause analysis on how the two static analysis tools (\ana and \crown) handle NPD bugs (\cc{Null} pointer and subsequent operations).}
\footnotesize
\begin{tabular}{l c c}
 \toprule
 \textbf{} & \textbf{\begin{tabular}[c]{@{}c@{}}C2Rust-analyze\end{tabular}} & \textbf{CROWN} \\
 \midrule
 \textbf{Strategy} & Refactor, runtime handles errors & Only refactor what is proven safe \\
 \textbf{\cc{NULL} Refactoring} & \cc{Option<\&mut T>} & \cc{*mut T} \\
 \textbf{\cc{NULL} Operation} & Runtime with panic on \cc{unwrap} & \cc{*T} (dereference) \\
 \bottomrule
\end{tabular}
\label{tab:crown}
\end{table}

\bheading{Precision-based type refactoring reverts to unsafe raw pointers when program complexity exceeds thresholds.}
The traditional refactoring tools (\ana and \crown) perform static analysis, which is used to infer whether types of C program meet the safety characteristics of Rust, before refactoring the entire C code. \ana identifies the permission (e.g. READ, WRITE) required for all memory operations, and then refactors the types for each set of permission. In contrast, \crown uses a precision-based ownership analysis to decide the type refactoring. This strategy consists of a conservative MIR-level ownership inference. When \crown encounters a pointer that it cannot classify as safe with high precision, it defaults to a raw pointer instead of a refactoring. In~\autoref{tab:crown}, it compares the different designs and strategies of \ana and \crown in the case of NPD bugs. The rationale behind this is to preserve type compatibility; however, it becomes the main cause of the mitigation failures of \crown. The result shows that \crown does not refactor the types in all our C programs into safe Rust types. After studying intermediate results of \crown, we find that while \crown can identify the allocated pointers (e.g., \cc{malloc}, \cc{calloc}) for ownership with high precision, precision is downgraded by the complexity of programs. \crown calculates complexity for each function based on the depth of pointer dereference, the depths of loops, and external function calls to avoid path explosion of the SAT solver. Therefore, \crown sets rules to decrease precision when it reaches the threshold. When precision reaches 0, the decision of refactoring will be returned to keeping as a raw pointer. For example, in the case of UAF bugs, \crown correctly identifies the pointer \cc{data} with ownership. However, since the depth of the dereference reaches 2 (\cc{*data.offset(0) = value;}), it triggers the condition of \cc{no\_box=true} to convert the \cc{Box} pointer back to a raw pointer.

\bheading{Compilation-priority rollback abandons safety enhancements when compilation errors persist.}
This strategy uses a rollback mechanism when compilation errors cannot be fixed. The rationale is to preserve the original compilation when the Rust code cannot be successfully refactored. \safer uses this strategy to fix all compilation errors in each function and performs a rollback if errors exist after 5 attempts of the self-debug process (5 is set in our experiment).
The rollback mechanism abandons all safety enhancements in the function and preserves the original bug semantics, which is responsible for most of the failures in mitigation of \safer.
In contrast, \fluo applies different granularity in the fixing of compilation errors. It focuses on one error in each attempt, which is different from \safer's strategy of whole function regeneration. Since the targeted error fix can make smaller code changes, it is simpler to preserve existing safety enhancements while fixing compilation errors. With this architectural choice between fixing granular errors and restoring the whole function, \fluo achieves greater security enhancement in bug mitigation than \safer.

\vspace{0.5em}
\begin{mdframed}[
  roundcorner=5pt,
  linewidth=0.8pt,
  linecolor=black,
  backgroundcolor=gray!20,
  innertopmargin=6pt,
  innerbottommargin=6pt,
  innerleftmargin=6pt,
  innerrightmargin=6pt
]
\textbf{Summary for RQ2.}
Syntax refactoring strategy merely refactors C bugs into Rust panics, while the conservative precision-based refactoring can only analyze simple programs.
While LLMs are better at refactoring due to semantic inference and error handling,
they will omit safety enhancements if necessary when prioritizing compilation success.
\end{mdframed}

%% file: RQ3.tex
\section{RQ3: \rqthree} \label{extra-issues}

In this section, we evaluate whether these refactoring tools can introduce additional Rust bugs that are not present in the original C programs, with the aim of answering RQ3.
As shown in \autoref{tab:overfit}, the evaluated tools can introduce up to \raf{77} additional bugs during the refactoring process.
Among refactoring tools, a significant variation in outcomes is observed: C2Rust-analyze leads to the most substantial increase in additional bugs, while FLUORINE results in the least.
We will summarize the cases of additional bugs introduced by different tools as follows.

\begin{table}[t]
\centering
\caption{Additional bug types from the angles of refactoring strategies. Categories combine two perspectives: the source of a bug (e.g., inheritance from C2Rust, constraint conflicts in LLM prompts) and its observable symptom (e.g., panic across FFI boundaries, panic breaking original semantics). We adopt this combined classification because individual tools exhibit patterns at different levels: static-analysis tools produce bugs identifiable by source, while LLM-driven tools produce bugs more reliably identifiable by symptom. The last row ($x/y$) shows the number of additional bugs and total programs.}
\footnotesize
\setlength{\tabcolsep}{4pt}
\begin{tabular}{lcccc}
\toprule
 & \multicolumn{3}{c}{\textbf{C2Rust-based}} & \textbf{Direct} \\
\cmidrule(lr){2-4} \cmidrule(lr){5-5}
\textbf{Additional Bug Type}
 & \textbf{\begin{tabular}[b]{@{}c@{}}C2Rust-\\analyze\end{tabular}}
 & \textbf{CROWN}
 & \textbf{\begin{tabular}[b]{@{}c@{}}C2Safer-\\Rust\end{tabular}}
 & \textbf{FLUORINE} \\
\midrule
Bug inheritance from C2Rust & 11 & 7 & 1 & 0 \\
Panic across FFI boundaries (UB) & 40 & 4 & 2 & 0 \\
Panic breaking original semantics & 1 & 0 & 1 & 1 \\
Constraint Conflicts to Violations & 0 & 0 & 0 & 6 \\
Non-deterministic LLM artifacts & 0 & 0 & 4 & 0 \\
\midrule
Total & 52 / 87 & 11 / 105 & 8 / 105 & 6 / 116 \\
\bottomrule
\end{tabular}
\label{tab:overfit}
\end{table}

\subsection{Methodology}
To expand the scope of broadening Rust bug detection, we employ Clippy and an LLM agent instructed by undefined behavior rules. Clippy serves as a static analyzer, relying on HIR code patterns to identify specific issues. Although Clippy enforces over 750 rules targeting common errors, the majority focus on syntax rather than directly addressing undefined behaviors. Consequently, we limit Clippy's use to 6 rules specifically aimed at detecting unsound type conversions and thread safety concerns. For the LLM agent, we utilize the Claude Sonnet 4 model and provide it with system instructions detailing undefined behaviors in Rust~\cite{ub}. To prevent hallucination, we integrate the ASan and Miri function tools to confirm bug presence. ASan and Miri together enhance bug detection coverage that might be missed by either tool alone. For instance, ASan isn't aimed at identifying certain undefined behaviors in Rust, while Miri lacks support for analyzing numerous C functions. Ultimately, when the function tools supply execution outputs, the LLM agent can determine if any additional runtime panic occurs in the refactored Rust code.

\subsection{Results}
In \autoref{tab:overfit}, we present the analysis of \raf{77} additional bugs introduced by the refactoring tools. The table reveals that refactoring based on C2Rust tend to be more prone to inheriting bugs from the translation phase involving C2Rust and encounters with panic across FFI boundaries. \ana exhibits a greater number of inherited bugs from C2Rust compared to the other two tools based on C2Rust, indicating that \crown and \safer perform type refactoring with higher success. On the other hand, \fluo completely sidesteps undefined behavior or panic across FFI interfaces but introduces distinct bug types stemming from constraint conflicts absent in the other three refactoring tools. There are 4 unclassified bugs, many of which are likely due to the non-deterministic code generation typical of LLMs.

\noindent\begin{minipage}[t]{0.48\textwidth}
    \begin{lstlisting}[
        basicstyle=\ttfamily\scriptsize,
        frame=lines,
        numbers=left,
        breaklines=true,
        showstringspaces=false,
        caption={Original C implementation with \cc{alloca} causing use-after-free bug.},
        label=lst:c_original
    ]
void bad()
{
    int * data;
    data = NULL;
    if(staticReturnsTrue())
    {
        data = (int *)ALLOCA(10);
    }
    {
        int source[10] = {0};
        size_t i;
        for (i = 0; i < 10; i++)
        {
            data[i] = source[i];
        }
        printIntLine(data[0]);
    }
}
    \end{lstlisting}
\end{minipage}%
\hspace{0.04\textwidth}
\begin{minipage}[t]{0.48\textwidth}
    \begin{lstlisting}[
        language=Rust,
        basicstyle=\ttfamily\scriptsize,
        frame=lines,
        numbers=left,
        breaklines=true,
        showstringspaces=false,
        caption={C2Rust refactoring mapping \cc{alloca} to \cc{vec::from\_elem} without considering scope.},
        label=lst:rust_transformed
    ]
pub unsafe extern "C" fn bad() {
    let mut data = 0 as *mut libc::c_int;
    data = 0 as *mut libc::c_int;
    if staticReturnsTrue() != 0 {
        let mut fresh0 = vec::from_elem(0, 10);
        data = fresh0.as_mut_ptr();
    }
    let mut source: [libc::c_int; 10] =
        [0, 0, 0, 0, 0, 0, 0, 0, 0, 0];
    let mut i: size_t = 0;
    i = 0;
    while i < 10 {
        *data.offset(i) = source[i];
        i = i.wrapping_add(1);
        i;}
    printIntLine(*data.offset(0));
}
    \end{lstlisting}
\end{minipage}

\subsection{Root Cause Analysis}
Following the bug detection phase, we conduct a root cause analysis to classify the additional Rust bugs. We subsequently outline two primary strategies accountable for the majority of introduced bugs: The first is C2Rust inheritance, which reflects the design decisions in the refactoring tools. The second involves safety reliance on panic, which originates from the syntax refactoring intended for safety improvements.

\bheading{C2Rust Inheritance propagates undefined behaviors from literal translation strategies.}
This refactoring strategy adopts C2Rust as the foundational transpilation step, inheriting its literal translation strategy. \ana, \crown, and \safer apply this strategy as the initial step to prioritize functional equivalence over idiomatic safety. This inheritance-based approach serves a critical purpose: C2Rust provides a mechanically sound starting point that preserves C semantics through unsafe Rust code, allowing subsequent safety analysis tools to operate on syntactically valid Rust code. However, when these specialized tools fail to successfully or only partially refactor C2Rust's output (e.g., see \crown's precision-based type refactoring or \safer's compilation-priority rollback in~\autoref{miti-fail}), they inherit and propagate the undefined behavior patterns introduced by C2Rust's translation heuristics. Specifically, based on our dataset, C2Rust applies error-prone strategies in the use of \cc{transmute}. The two most common patterns are the \textit{immutable to mutable} refactoring (\cc{transmute::<\&[u8], \&mut [libc::c\_char]>}) and \textit{misalignment} refactoring (\cc{transmute::<\&[u8], \&[libc::c\_int]>}), which are undefined behaviors and can cause data races or run-time panic. More critically, we find that C2Rust applies a scope-agnostic refactoring: it translates C's function without taking lifetime scope into consideration, creating use-after-free bug. In~\autoref{lst:c_original}, original C code calls \cc{alloca} in the code block of the if-condition. C2Rust directly maps \cc{alloca} to \cc{vec::from\_elem} in the same code block but is unaware that the memory will be released at the end of the code block in~\autoref{lst:rust_transformed}. This observation highlights that: First, C2Rust also applies syntax refactoring without semantic understanding cause an additional Rust bug. Second, while using C2Rust as the initial step enables a sophisticated safety analysis, it also introduces Rust-specific memory security violations\footnote{Note that the recent release of C2Rust (0.21.0) addresses several of the \cc{transmute} patterns and the \cc{alloca} issue discussed in this section. The underlying design pattern of literal translation without semantic understanding, however, remains and continues to be inherited by downstream refactoring tools.}.

\bheading{Panic safety on FFI boundaries and Semantic integrity.}
This refactoring strategy is derived from syntax refactoring and rely on run-time panic to improve safety. As described in~\autoref{miti-fail}, panic will cause the program to stop and require a more elegant error handling mechanism. While studying potentially additional Rust bugs, we find that reliance on panic can also introduce an undefined behavior in Rust.
Especially in the FFI scenario where the external C functions remain, panic across the FFI boundaries is defined as an undefined behavior\footnote{A subset of programs may appear both in \autoref{tab:repair}'s Table (runtime panic is counted as a mitigation strategy for the original C bug under RQ2) and in \autoref{tab:overfit} (panic crossing a FFI boundary is counted as a newly introduced bug under RQ3). The two categorizations answer distinct questions: whether the original vulnerability is no longer reachable, and whether a new vulnerability has been introduced. We do not view this overlap as double-counting: the same panic can simultaneously remove an NPD-reachable dereference and introduce a stack-unwinding hazard when the refactored function is invoked from C.}. The reason is that when a Rust function panics, it typically performs stack unwinding to call destructors for all live variables. If the unwinding process continues into the C code's stack frames, it will cause to unpredictable results due to ABI mismatches.
\ana, \crown, and \safer demonstrate the risk of this undefined behavior. All refactored Rust functions provided by \ana and \crown are declared to be \cc{pub unsafe extern "C" fn}, which can be called by external C functions. When \safer performs a rollback to the original code, it also provides the same types of function. The rationale behind this is incremental refactoring. For example, there are two C functions, one as the caller and the other as the callee. While the callee function is refactored into safe Rust by the refactoring tools, developers can keep the caller function in C and seamlessly invoke the callee function written in Rust. This strategy is especially practical in the real-world C code base, such as Linux Kernel~\cite{Rustlinuxkernel}. Although these code bases are manually rewritten, developers also use the same strategy to incrementally refactor C code into Rust. Even though the original C bug is mitigated with the panic mechanism in safe Rust, it can trigger the additional undefined behavior when involved in the workspace of incremental rewriting with refactoring tools. In addition, there is a distinct source of panic that is categorized as \textit{panic breaking original semantics} in~\autoref{tab:overfit}, which arises from API-level mapping: \safer maps C operations to their closest Rust safe equivalents (e.g., \cc{strcpy} to \cc{copy\_from\_slice}), which panic on inputs the original C would accept. Unlike FFI-boundary panics, this form changes observable behavior even when the caller is Rust.


\bheading{Constraint-based Prompts Design missing Conflict Resolution cause to Safety Violations.}
After studying the prompt design of LLM-based approaches, we find that they could simultaneously enforce multiple conflicting objectives: semantic preservation, safety constraints (avoiding unsafe Rust patterns), and compilation requirements. The rationale behind this approach is to leverage the LLM's broad knowledge to automatically resolve complex refactoring challenges while satisfying requirements for practical constraints for code correctness. However, this strategy can fail when these objectives become mutually incompatible because the tool lacks a systematic conflict resolution mechanism. \fluo is one of the LLM-based tools that uses this strategy and demonstrates the errors. The tool has no mechanism to recognize this conflict or choose an appropriate resolution strategy. Instead, \fluo's implementation only checks constraint violations after the LLM has already generated code by performing a simple text pattern matching, such as searching for \cc{unsafe} keywords in responses to ``safe Rust'' prompts. This post hoc checking approach means that violations are measured, but never prevented or corrected. The tool prioritizes compilation success, but abandons safety enhancements. While \safer also prioritizes the compilation success, it will roll back to the output of C2Rust instead of the code generation of LLM. We also find that the prompt of \fluo explicitly lists six constraints including ``uses safe rust'' and ``Don't use raw pointers'', yet the LLM completely ignores these requirements (see example in~\autoref{app:promptdesign}).


\vspace{0.5em}
\begin{mdframed}[
  roundcorner=5pt,
  linewidth=0.8pt,
  linecolor=black,
  backgroundcolor=gray!20,
  innertopmargin=6pt,
  innerbottommargin=6pt,
  innerleftmargin=6pt,
  innerrightmargin=6pt
]
\textbf{Summary for RQ3.}
Refactoring introduces new bugs due to fundamental design trade-offs that prioritize syntax conversion over semantic preservation.
The semantics can be broken during refactoring due to inconsistent APIs between C/Rust libraries and additional constraints embedded in LLM prompts.
Rust's panic safety unintentionally leads to undefined behavior propagation across FFI boundaries.
\end{mdframed}

%% file: Discussion.tex
\section{Discussion}
\label{sec:discuss}

The scope of our claims is shaped by the current state of C-to-Rust refactoring tools and by the security-focused questions we set out to answer. We discuss below how our methodology aligns with that scope.

\bheading{Coverage of bug detection.}
No single bug-detection tool covers all classes of memory-safety bugs in refactored Rust code. ASan does not flag Rust-specific undefined behaviors; Miri cannot execute across FFI boundaries; Clippy focuses on syntactic patterns. We combine all three with an LLM agent guided by Rust's undefined-behavior specification precisely because their coverage is complementary. The numbers we report should therefore be read as a lower bound on the bugs that automated refactoring introduces; the attack surface is at least as large as what our measurements reveal.

\bheading{Role of the LLM agent.}
Claude Sonnet 4 serves two auxiliary roles in our pipeline: repairing trivial compilation errors in RQ1 and assisting UB detection in RQ3. Neither role is focus of our paper's contributions. The security-relevant findings such as precision-based rollback in CROWN, compilation-priority rollback in C2SaferRust, C2Rust inheritance across three of the four tools, and prompt-level constraint conflicts in FLUORINE, are properties of the refactoring tools themselves, revealed by our evaluation but not produced by it. A different LLM might resolve a different subset of compilation errors or flag bugs with a different recall rate, but the design-level observations are independent of the choice of auxiliary model.

\bheading{Scope of the evaluation.}
We evaluate whether automated refactoring preserves or undermines memory security. Whether the refactored code is semantically equivalent to the original C is a separate question, studied by the refactoring tools themselves through differential testing and formal verification~\cite{fluorine,yang2024vert,shetty2024syzygy}. Our findings are orthogonal to that line of work: a program that introduces undefined behavior at an FFI boundary is unsafe regardless of whether it preserves the original semantics.

\bheading{Dataset and tool selection.}
Our evaluation uses 116 programs from the NIST Juliet Test Suite, covering the CWE Top~25 memory-safety categories (NPD, UAF, DF, BOF, TC), and four state-of-the-art refactoring tools spanning the two architectural axes of the current landscape: static-analysis versus LLM-based, and C2Rust-based versus direct translation. The Juliet-based evaluation is the most faithful measurement currently available: we verified in separate experiments that the tools' frontends fail to parse real-world codebases such as OpenSSL~\cite{cve201820997} and radare2~\cite{radare}, so any real-world evaluation today would measure frontend coverage rather than refactoring behavior. The failure modes we identify are tied to design decisions rather than specific implementations, and new tools entering the field are likely to encounter analogous trade-offs.

%% file: RelatedWork.tex
\section{Related Work}

\bheading{C-to-Rust Translations and Refactoring.}
In recent years, more automatic techniques are developed for transforming C code into Rust, driven by the desire for Rust's safety guarantees. C2Rust~\cite{c2rust} is one of the most widely adopted tools for syntactic translation from C to unsafe Rust. However, to generate more idiomatic Rust code, recent works have proposed refactoring strategies that incorporate static analysis and Large Language Models (LLMs)~\cite{laertes,crown,c2rustanalyze,nitin2025c2saferrust,hong2024tag,hong2024don}. Other approaches directly translate C code using LLM-guided refactoring~\cite{yang2024vert,fluorine,shetty2024syzygy}. Although these studies compare the effectiveness of different tools, their evaluations typically emphasize syntactic refactoring quality and the use of safe Rust pointers, while often overlooking the security implications related to bug mitigation.
Ruishi~et al.~\cite{userstudy} analyzed translation strategies from user points of view, yet did not focus on bug fixing effectiveness in depth.
Pan~et al.~\cite{lostranslation} investigated the ability of general LLMs and code LLMs for code generation across pairs of different programming languages, including C and Rust. However, their study is limited in C2Rust, which mainly generates unsafe Rust code, but does not include safety refactoring.
Our study fills these gaps by conducting a comprehensive analysis of security implications on refactoring strategies, which are applied by modern tools that focus on transforming C code into safe Rust code.

\bheading{Safety Issues of Unsafe Rust.}
Extensive research has examined how the use of \cc{unsafe} can undermine the Rust's safety guarantees~\cite{Qin2020ReplicationPF,Evans2020IsRU,Xu2020MemorySafetyCC,zhang2022towards,papaevripides2021exploiting,mergendahl2022cross,rivera2021keeping,kirth2022pkru,unsafeachilles}.
Xu~et al.~\cite{Xu2020MemorySafetyCC} analyzed hundreds of memory security issues and considered \cc{unsafe} code the root cause.
Cui~et al. further categorized the safety requirements for unsafe functions that are strongly correlated with numerous real-world CVEs.
While some studies focus on essential uses of \cc{unsafe} in low-level and embedded systems~\cite{embeded,rust4linux}, others propose mechanisms to mitigate risks arising from \cc{unsafe} code~\cite{rivera2021keeping,kirth2022pkru}.
Static analysis tools such as Rudra~\cite{Yechan2021Rudra} and MirChecker~\cite{Zhuohua2021MirChecker} have been developed to detect memory security bugs introduced by unsafe code.
All these works highlight the security implications of \cc{unsafe} usage and the importance of promoting safe Rust development.
Compared to previous works, our study focuses on a new surface: the potential security risks introduced by automatic C-to-Rust refactoring, which often retain or generate \cc{unsafe} code.
We explore how these refactoring can inadvertently compromise safety.

%% file: Conclusion.tex
\section{Conclusion}

This paper presents a comprehensive study of automatic C-to-Rust refactoring tools, evaluating four state-of-the-art tools on 116 buggy C programs from the NIST Juliet Test Suite. Our results indicate that these tools are unable to correctly compile \raf{342} Rust programs, not effectively enhance the safety of \raf{177} Rust programs, and may introduce as many as \raf{77} new bugs through a refactoring tool. The root causes include compilation failures from type mismatches, mitigation failures from syntax refactoring without semantic understanding, and additional bugs caused by C2Rust and further refactoring. Our results demonstrate that automatic C-to-Rust refactoring does not yet deliver the memory security, even when the tools successfully produce idiomatic Rust code. Until refactoring tools treat security as a design constraint rather than an emergent property of reducing \cc{unsafe} code, developers should continue auditing the output Rust program.

\section{Data Availability}
Our dataset and executable scripts are available at: \url{https://anonymous.4open.science/r/C-to-Rust-Fallacy-E752/}

%% file: Appendix.tex
\section{Bug Type Dataset}
\label{app:bugtype}

\begin{table}[]
    \centering
    \caption{The memory-safety bug dataset.}
    \label{tab:dataset}
    \footnotesize
    \begin{tabular}{c c c}
    \toprule
    \textbf{Bug Type} & \textbf{CWE} & \textbf{Count}  \\
    \midrule
    Null Pointer Deref & \makecell{CWE690\\CWE476} & \makecell{22\\18} \\
    \midrule
    Use After Free & CWE416 & 14 \\
    \midrule
    Double Free & CWE415 & 12  \\
    \midrule
    Buffer Overflow & \makecell{CWE121\\CWE122} & \makecell{24\\22}  \\
    \midrule
    Type Confusion & CWE843 & 4 \\
    \bottomrule
    \end{tabular}
\end{table}

\autoref{tab:dataset} lists the CWE-to-count distribution of our 116-program dataset. Each CWE category is represented by multiple programs to cover different bug-triggering patterns, following the classification of the NIST Juliet Test Suite.

\section{Control-Flow and Data Type Features of Dataset}
\label{app:datafeature}

\begin{table}[]
        \centering
        \caption{Buggy C programs with different control flow features. Note that a single C program can have multiple control flow features, so the total percentage values exceed 100\%.}
        \label{tab:control-flow-stats}
        \footnotesize
        \begin{tabular}{l c c}
            \toprule
            \textbf{Pattern} & \textbf{Count} & \textbf{Percentage} \\
        \midrule
        Cross-function call & 32 & 27.6\% \\
        Conditional branch (if/else) & 95 & 81.9\% \\
        Loop (for/while) & 41 & 35.3\% \\
        Switch statement & 2 & 1.7\% \\
        \makecell[l]{Global variable} & 22 & 19.0\% \\
        \bottomrule
        \end{tabular}
\end{table}

\begin{table}[]
        \centering
        \caption{Buggy C programs with different pointer types. A single C program also has multiple data types involved so that total percentage exceeds 100\%.}
        \label{tab:data-type-stats}
        \footnotesize
        \begin{tabular}{l c c}
            \toprule
            \textbf{Pattern} & \textbf{Count} & \textbf{Percentage} \\
        \midrule
        Stack allocation & 45 & 38.8\% \\
        Heap allocation & 75 & 64.7\% \\
        Struct usage & 10 & 8.6\% \\
        Pointer alias & 99 & 85.3\% \\
        Function pointer & 5 & 4.3\% \\
        Void pointer & 13 & 11.2\% \\
        \bottomrule
        \end{tabular}
\end{table}

Beyond the CWE categorization, we characterize our dataset along two dimensions that influence how refactoring tools behave: control-flow features (\autoref{tab:control-flow-stats}) and pointer-type features (\autoref{tab:data-type-stats}). Because a single program typically exhibits multiple features, percentages across categories sum to more than 100\%. This breakdown allows readers to assess how representative our dataset is of the code patterns targeted by each refactoring strategy.

\section{Constraint-based Prompt Design}
\label{app:promptdesign}

\begin{lstlisting}[
    language=rust,
    style=lst,
    caption={Line 1-6 show 6 constraints enforced in the prompt for LLM. The vulnerable function starts from line 7. While \cc{CString::from\_raw} relies on automatic memory management on line 15, LLM still manually releases the memory with \cc{dealloc} on line 17, which leads to double free bug.},
    label=conflict,
    mathescape=true,
    basicstyle=\ttfamily\scriptsize,
    belowskip=-0.5em
]
// 1. Use the same function name, same argument and return types.
// 2. Make sure it includes all imports, uses safe rust, and compiles.
// 3. Don't use raw pointers.
// 4. Use box pointer whenever possible. Box pointers are preferable to other alternatives.
// 5. Try not to use Traits if possible. I would not like to have Traits in Rust code.
// 6. Try not to use Generics if possible.
fn sink(data: Box<[u16]>) {
    unsafe {
        let data_len = data.len();
        let layout = Layout::array::<u16>(data_len.wrapping_add(1)).unwrap();
        let dest = alloc(layout) as *mut 16;
        if dest.is_null { exit(-1); }
        ptr::copy_nonoverlapping(data.as_ptr(), dest, data_len);
        *dest.add(data_len) = 0; // Null-terminate the wide string
        let c_str = CString::from_raw(dest as *mut i8);
        print_line(c_str.to_str().unwrap());
        dealloc(dest as *mut u8, layout);
    }
}
\end{lstlisting}

\section{Recommendations}
\label{app:recom}
Based on our comprehensive evaluation of state-of-the-art C-to-Rust refactoring tools across compilation correctness, bug mitigation effectiveness, and security implications, we provide actionable recommendations for future tool developers.

\bheading{Adopting hybrid approaches for compilation correctness.}
Our findings reveal that pure static analysis approaches suffer from limitations in handling the semantic gap between C and Rust, while pure LLM approaches risk semantic divergence and unpredictable behaviors. We recommend developing hybrid architectures that combine the predictability of static analysis with the adaptability of LLMs. Static analysis should be used to establish semantic constraints and define refactoring patterns, while LLMs should handle complex syntactic repairs and pattern recognition within these constraints. This approach can address type mismatch errors (E0308) from type refactoring and the unstable feature dependencies (E0658) that compromise production deployment. Specifically, developers should implement hybrid approaches where error handling mechanism is specified in prompts and using static analysis to provide explicit program workflow to narrow the scope of LLM-generated changes, preventing overly aggressive optimizations that alter program semantics.

\bheading{Designing FFI-aware refactoring architectures.}
Current tools attempt to eliminate the FFI boundaries or ignore their implications, leading to unsafe refactorings and undefined behavior related to panic. We recommend that future approaches consider FFI as a necessary component of real-world C-to-Rust migration and design refactoring strategies for hybrid language environments. This includes implementing ownership tracking across language boundaries, developing safe wrappers for cross-language memory management, and ensuring that panic unwinding never crosses FFI boundaries. The research field of semantics comparison between C and Rust should also be explored. Current approaches of semantic verification between C and Rust are still restricted in testing and type equivalence. The FFI-aware architecture should identify the hidden invariants (e.g., lifetime, aliasing, initialization, provenance) that C's compiler never records. Furthermore, integration with FFI-aware bug detection tools should be implemented to provide iterative refinement and validation of cross-language interactions~\cite{10.5555/3620237.3620626,10.1145/3625275.3625397,ffichecker,hu2022crust}.

\bheading{Establishing precision-aware and incremental refactoring mechanisms.}
The failures of precision-based type refactoring observed in CROWN highlight the need for more sophisticated approaches to handling complicated code patterns. Rather than reverting to unsafe code when precision thresholds are exceeded, tools should implement incremental refactoring strategies that make partial safety improvements while maintaining compilation correctness. This involves developing algorithms that can identify which portions of complex code can be safely refactored and which require manual intervention. Tools should provide transparency about their decision-making process and provide developers with clear guidance on how to address limitations. Additionally, incremental refactoring mechanisms should be designed with proper type conversion handling to avoid the compilation errors that arise from mixing refactored and legacy code.

\bheading{Prioritizing error recovery and fixing strategies.}
Our evaluation demonstrates that compilation-priority rollback mechanisms that abandon entire functions when faced with errors are counterproductive to improving safety. We recommend implementing error-fixing approaches that target specific compilation issues while preserving existing safety improvements. This involves developing error classification systems that can distinguish between critical errors requiring rollback and minor issues that can be fixed locally. Tools should maintain intermediate safety enhancements even when faced with compilation challenges, and provide developers with detailed feedback about which refactoring succeeded and which require attention. Furthermore, error recovery mechanisms should include semantic preservation checks to ensure that fixes do not inadvertently introduce new bugs or alter program behavior.

\bheading{Integrating comprehensive validation and testing frameworks.}
The introduction of additional bugs during refactoring, particularly at FFI boundaries and through constraint conflicts, requires robust validation mechanisms. Future tools should integrate multiple bug detection approaches, including static analysis (Clippy), dynamic analysis (ASan, Miri), and semantic verification techniques. These validation frameworks should be designed to work effectively with partially refactored code and provide meaningful feedback about both safety improvements and potential regressions. Tools should also implement differential testing capabilities to verify semantic equivalence between original C code and refactored Rust code, adapting these mechanisms to work with isolated code fragments typical of incremental refactoring scenarios.